\documentstyle[preprint,tighten,eqsecnum,aps,floats,psfig,epsfig]{revtex}

\begin{document}
\draft
\title{
Multicritical phenomena in O($n_1$)$\oplus$O($n_2$)-symmetric theories.
}
\author{Pasquale Calabrese,$^1$ Andrea Pelissetto,$^2$ 
Ettore Vicari$^3$ }
\address{$^1$ Scuola Normale Superiore and  INFN, P.za Cavalieri 7,
I-56126 Pisa, Italy}
\address{$^2$ Dip. Fisica dell'Universit\`a di Roma ``La Sapienza" \\
and INFN, P.le Moro 2, I-00185 Roma, Italy}
\address{$^3$
Dip. Fisica dell'Universit\`a di Pisa
and INFN, 
V. Buonarroti 2, I-56127 Pisa, Italy}
\address{
{\bf e-mail: \rm 
{\tt calabres@df.unipi.it},
{\tt Andrea.Pelissetto@roma1.infn.it},
{\tt vicari@df.unipi.it}
}}

\date{\today}

\maketitle

\begin{abstract}
We study the multicritical behavior arising from the
competition of two distinct types of ordering characterized by O($n$) 
symmetries. For this purpose, we consider the 
renormalization-group flow for the most general 
O($n_1$)$\oplus$O($n_2$)-symmetric
Landau-Ginzburg-Wilson Hamiltonian involving two fields $\phi_1$ and $\phi_2$
with $n_1$ and $n_2$ components respectively.
In particular, we determine in which cases,
approaching the multicritical point, one may observe
the asymptotic enlargement of the symmetry to O($N$) with $N=n_1+n_2$.

By performing a five-loop $\epsilon$-expansion computation we determine 
the fixed points and their stability.
It turns out that for $N=n_1+n_2\ge 3$ 
the O($N$)-symmetric fixed point is unstable.
For $N=3$, the multicritical behavior is described by the biconal fixed point 
with critical exponents that are very close to the Heisenberg ones.
For $N\ge 4$ and any $n_1,n_2$ the critical behavior is controlled 
by the tetracritical decoupled fixed point.

We discuss the relevance of these results for some physically interesting
systems, in particular for anisotropic antiferromagnets in the presence 
of a magnetic field and for high-$T_c$ superconductors.
Concerning the SO(5) theory of superconductivity, we show
that the bicritical O(5) fixed point is unstable with a significant crossover 
exponent, $\phi_{4,4}\approx 0.15$; this implies that
the O(5) symmetry is not effectively realized at the 
point where the antiferromagnetic and superconducting transition 
lines meet. The multicritical behavior is either governed by the tetracritical
decoupled fixed point or is of first-order type if the system is outside 
its attraction domain.

\end{abstract}



\section{Introduction.}
\label{intro}

The competition of distinct types of ordering gives rise to multicritical
behavior. More specifically, a multicritical point (MCP) is observed at the 
intersection of two critical lines characterized by different order parameters.
MCP's arise in several physical contexts.  The phase diagram of anisotropic 
antiferromagnets in  a uniform magnetic 
field $H_\parallel$ parallel to the anisotropy axis presents
two critical lines in the temperature-$H_\parallel$ plane, 
belonging to the $XY$ and Ising universality 
classes, that meet at a MCP \cite{FN-74,NKF-74,KNF-76}.
A MCP is also observed in $^4$He. It arises from the competition of
crystalline and superfluid ordering in the
temperature-pressure phase diagram \cite{LF-72}.
MCP's are also expected 
in the temperature-doping phase diagram of
high-$T_c$ superconductors.
Within the SO(5) theory \cite{Zhang-97,ZHAHA-99}  
of high-$T_c$ superconductivity, it has been speculated that 
the antiferromagnetic and superconducting transition lines meet
at a MCP in the temperature-doping phase diagram,
which is bicritical and shows an effective enlarged O(5) symmetry.
On the other hand, the recent experimental evidence of
a coexistence region between the antiferromagnetic and
superconducting phases is suggestive of a tetracritical behavior \cite{ZDS-02}.
A MCP should also appear in the temperature baryon-chemical-potential 
phase diagram of hadronic matter, within the strong-interaction theory
with two massless quarks \cite{CCBW-01,CCBW-footnote}.

\begin{figure}[tb]
\centerline{\psfig{width=8truecm,angle=0,file=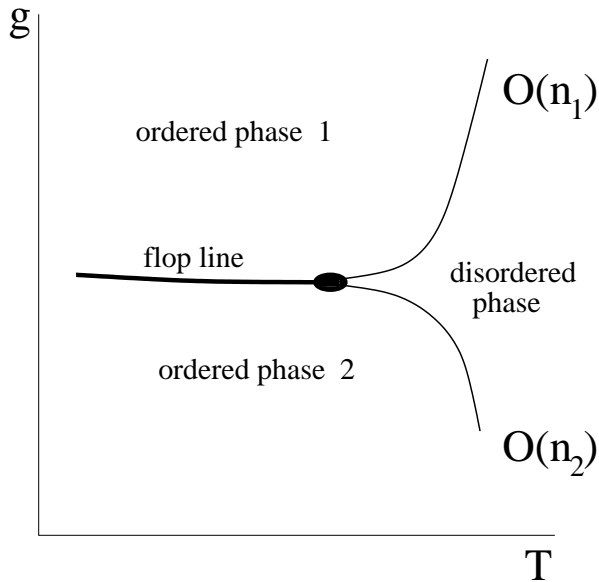}}
\vspace{2mm}
\caption{
Phase diagram in the plane $T$-$g$ presenting
a bicritical point. Here, $T$ is the temperature and $g$ 
a second relevant parameter.
The thick line (``flop line")
represents a first-order transition.
}
\label{bicr}
\end{figure}

\begin{figure*}[tb]
\centerline{\psfig{width=8truecm,angle=0,file=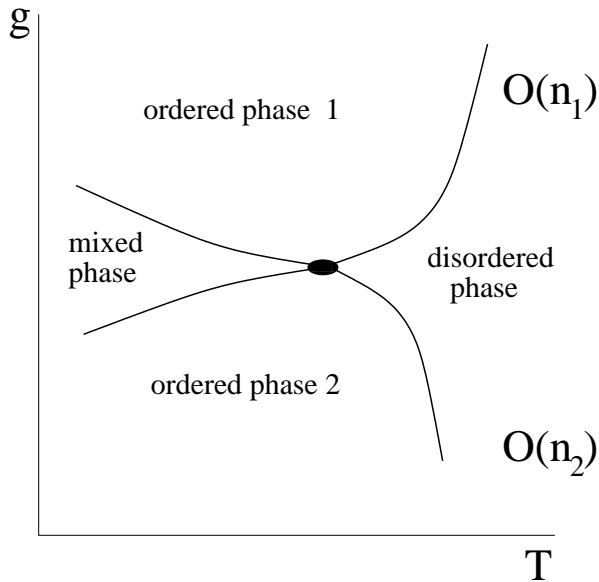}}
\caption{Phase diagram with a tetracritical point.
}
\label{tetra}
\end{figure*}

\begin{figure*}[tb]
\centerline{\psfig{width=8truecm,angle=0,file=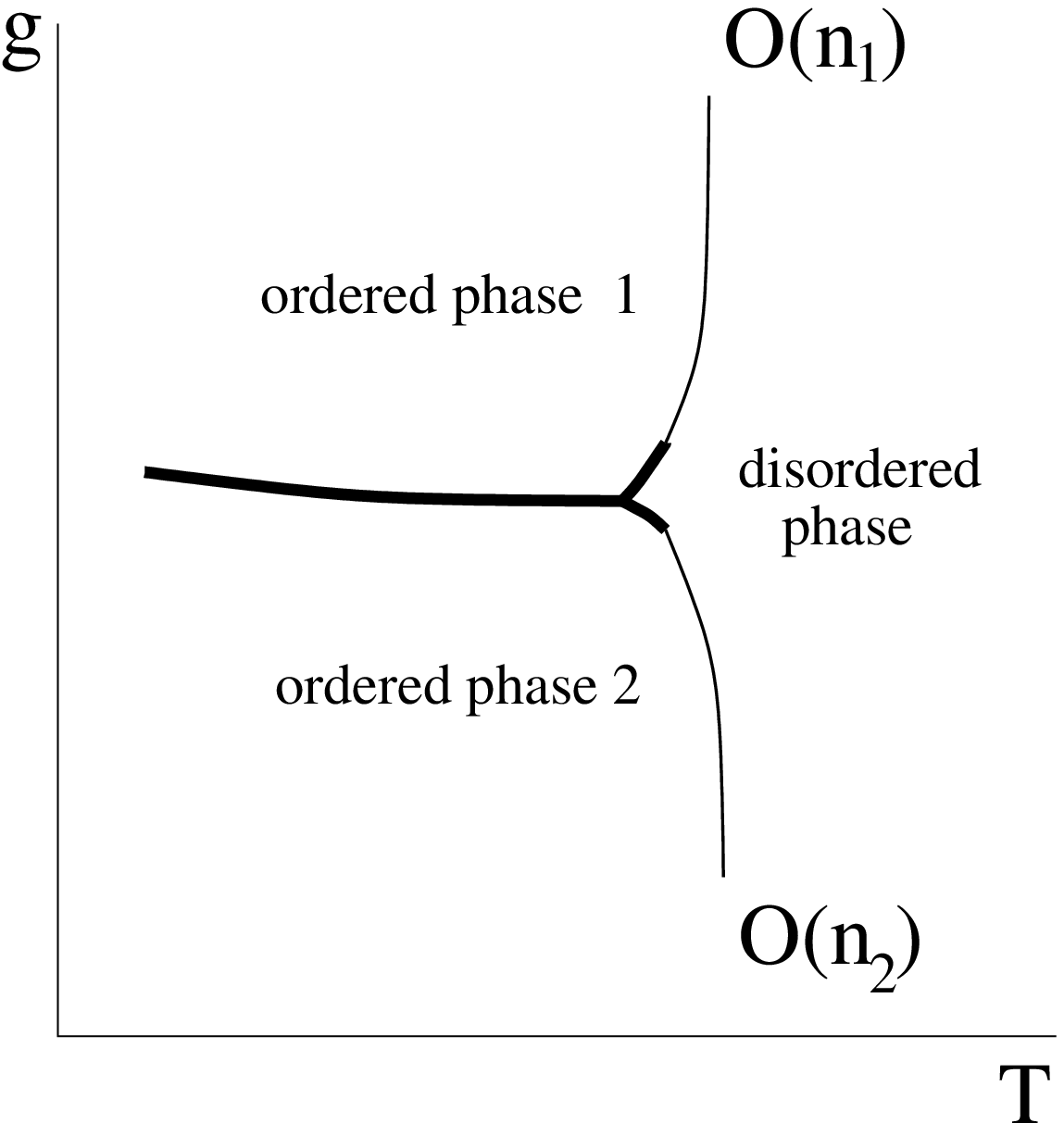}}
\caption{Phase diagram with a first-order MCP.
The thick lines represent first-order transitions.
}
\label{tricr}
\end{figure*}

Different phase diagrams have been observed close to a MCP.
If the transition at the MCP is continuous, one may observe 
either a bicritical or a tetracritical behavior.
A bicritical behavior is characterized by the presence
of a first-order line that starts at the MCP and separates
the two different ordered low-temperature phases, see Fig.~\ref{bicr}.
In the tetracritical case, there exists a mixed low-temperature phase  
in which both types of ordering coexist and which is bounded
by two critical lines meeting at the MCP, see Fig.~\ref{tetra}.
It is also possible that the transition at the MCP is of first order.
A possible phase diagram is sketched in Fig.~\ref{tricr}. In this 
case the two first-order lines, which start at the MCP and separate
the disordered phase from the ordered phases, end in tricritical points
and then continue as critical lines.

If the order parameters have respectively $n_1$ and $n_2$ components 
and the interactions are invariant under O($n_1$) and O($n_2$),
the critical behavior at the MCP
can be studied by starting from the most general
Landau-Ginzburg-Wilson (LGW) Hamiltonian 
that is symmetric under O($n_1$)$\oplus$O($n_2$) transformations
and contains up to quartic terms \cite{NKF-74}: 
\begin{eqnarray}
{\cal H} = \int d^d x \left\{ 
\case{1}{2} \Bigl[ ( \partial_\mu \phi_1)^2  + (
\partial_\mu \phi_2)^2\Bigr] + 
\case{1}{2} \Bigl( r_1 \phi_1^2  + r_2 \phi_2^2 \Bigr)  
+ \case{1}{4!} \Bigl[ u_1 (\phi_1^2)^2 + u_2 (\phi_2^2)^2 + 
                           2 w \phi_1^2\phi_2^2 \Bigr] \right\}.  
\label{bicrHH} 
\end{eqnarray}
Here, the two fields $\phi_1$ and $\phi_2$ have $n_1$ and $n_2$ components
respectively.
The critical behavior at the MCP is determined 
by the stable fixed point (FP)
of the renormalization-group (RG) flow when both $r_1$ and $r_2$ 
are tuned to their critical value.
An interesting possibility is that the stable FP has O($N$) symmetry, 
$N\equiv n_1 + n_2$, so that the symmetry gets effectively enlarged when 
approaching the MCP.
This picture has been 
put forward for the multicritical behavior of  anisotropic
antiferromagnets in an external magnetic field \cite{NKF-74,KNF-76},
for systems with quadratic and cubic anisotropy 
\cite{BA-75,DF-77,Corliss-etal-81},
and for high-$T_c$ superconductors \cite{Zhang-97,ZHAHA-99,HZ-00,Hu-01}.

The phase diagram of the model with
Hamiltonian (\ref{bicrHH}) has been investigated
within the mean-field approximation in Ref. \cite{LF-72}
(see also Ref.~\cite{KAE-01}).
This analysis predicts the existence of a bicritical or tetracritical
point, as observed experimentally. The nature of the MCP depends
on the sign of the quantity $\Delta=u_1 u_2 - w^2$, 
which is relevant in the study of 
the stability domain of the Hamiltonian (\ref{bicrHH}).
If $\Delta > 0$ the MCP is tetracritical as in Fig.~\ref{tetra}, 
while for $\Delta<0$  it is bicritical, as in Fig.~\ref{bicr}. 

The critical behavior of the model has been investigated
in the framework of the $\epsilon$ expansion
\cite{NKF-74,KNF-76}. A low-order calculation \cite{NKF-74,KNF-76} 
shows that the isotropic O($N$)-symmetric FP ($N\equiv n_1 + n_2$)
is stable for $N<N_c=4-2 \epsilon + O(\epsilon^2)$.
With increasing $N$, a new FP named biconal FP (BFP),
which has only O($n_1$)$\oplus$O($n_2$) symmetry, becomes stable. 
Finally, for large $N$, the decoupled FP (DFP)
is the stable FP. In this case, the two order parameters are 
effectively uncoupled at the MCP.
The extension of these $O(\epsilon)$ results to three
dimensions suggests that 
for $n_1=1$ and $n_2=2$, the case
relevant for anisotropic antiferromagnets,
the MCP belongs to  the O(3) universality class,
while  for $n_1=2$ and  $n_2=3$,
of relevance for the SO(5) theory of high-$T_c$ superconductivity,
the stable FP is the BFP.
The $O(\epsilon)$ computations provide useful indications on the 
RG flow in three  dimensions, but a controlled 
extrapolation to $\epsilon=1$ requires much longer series
and an accurate resummation exploiting their Borel summability.
As we shall see, the above-reported hypotheses 
on the three-dimensional systems with $n_1=1,\,n_2=2$ and $n_1=2,\,n_2=3$
will be both contradicted by a higher-order analysis.

The stability properties of the DFP can be  established
using nonperturbative arguments 
\cite{Aharony-76,GT-83,Aharony-02,Aharony-02-2}, 
which allow us to compute 
the RG dimension  $y_w$ of the operator $w\phi_1^2\phi_2^2$ at the DFP.
The stability of the DFP depends on the sign of $y_w$:
if $y_w<0$, the DFP is stable.
It turns out that in three dimensions $y_w>0$ for $N\le 3$,
and $y_w<0$ for $N\ge 4$ for any $n_1$ and $n_2$, showing 
that the DFP is stable for $N\ge 4$.
We should note that the stability of the DFP does not allow us to exclude
the existence of another stable FP.
This possibility, which is usually considered rather unlikely 
\cite{Aharony-02}, has been put forward \cite{Hu-02}
to explain the Monte Carlo results of  Refs. \cite{Hu-01,Hu-02}, 
which apparently support the stability of a multicritical O(5) FP.

The phase diagram of the model (\ref{bicrHH}) was studied in 
Refs.~\cite{BA-75,DF-77,Fisher-75}.
The DFP is expected to be generically tetracritical: indeed, in this case 
the MCP should correspond to a generic intersection of the two critical 
lines with O($n_1$) and O($n_2$) symmetry. The stable O($N$) FP---as 
we shall see, this is the case only for $N=2$---can be either bicritical 
or tetracritical. The possibility of two different phase diagrams
for the same FP is due to the presence of a dangerously irrelevant 
operator \cite{BA-75,DF-77}. Little is known for the BFP, although
a phenomenological extension of the mean-field 
arguments would predict a tetracritical behavior \cite{KNF-76}.
When the initial parameters of the Hamiltonian
are not in the attraction domain
of the stable FP, the transition between the disordered
and ordered phases should be of first order in the neighborhood of the MCP
\cite{HL-74,NT-75,MK-76}. However, the transition 
along the critical lines may become
continuous sufficiently far from the MCP
\cite{DMF-77,MN-00}. A possible phase diagram is sketched
in Fig.~\ref{tricr}. 

In this paper we extend  
the analysis of the multicritical RG flow to $O(\epsilon^5)$.
The stability of the O($N$) FP is also discussed in the framework
of fixed-dimension expansion in three dimensions, 
for which six-loop series have been computed.
These calculations allow us to obtain a rather conclusive picture
of the multicritical RG flow in three-dimensional systems.
In particular, the O($N$) FP is stable only for $N=2$. 
Therefore, the symmetry enlargement occurs only when the competing order 
parameters have Ising symmetry.  For $N\ge 3$, the O($N$) FP is unstable and
therefore the enlargement of the symmetry to O($N$) at the MCP 
requires an additional tuning of the parameters: beside tuning
$r_1$ and $r_2$, a third parameter must be properly fixed to 
decouple the additional relevant interaction. The crossover exponent
associated with this RG instability is 
$\phi_{4,4}\approx 0.01$ for $N=3$, 
$\phi_{4,4}\approx 0.08$ for $N=4$, 
$\phi_{4,4}\approx 0.15$ for $N=5$, and $\phi_{4,4}\rightarrow 1$
for $N\rightarrow \infty$.
For $N=3$ the stable FP is the BFP. 
The critical exponents are however very close
to the Heisenberg ones, so that distinguishing experimentally the O(3) FP 
and the BFP is a very hard task, taking also into account the very small
crossover exponent governing the unstable flow from the O(3) FP.
The case $N=5$, $n_1=2$, $n_2=3$ is relevant for the 
SO(5) theory \cite{Zhang-97,ZHAHA-99} of high-$T_c$ superconductors,
which proposes a description in terms of a three-component
antiferromagnetic order parameter and 
a $d$-wave superconducting order parameter with U(1) symmetry,
with an approximate O(5) symmetry.
For $N=5$ the only stable FP is the DFP which predicts, if the transition is
continuous, a tetracritical behavior.
This may explain a number of recent experiments, see, e.g., 
Refs.~\cite{ZDS-02,Khay-etal-02,Lee-etal-99,Katano-etal-00,%
Miller-etal-02,Lake-etal-01,Aeppli-etal-97},
that provided evidence of a coexistence region
of the antiferromagnetic and superconducting phases.
The O(5) FP is unstable with a crossover exponent $\phi_{4,4}\approx 0.15$,
which, although rather small, is nonetheless sufficiently large  
not to exclude the possibility of observing 
the RG flow towards the eventual asymptotic behavior 
for reasonable values of the reduced temperature \cite{ahooo},
even in systems with a moderately small breaking of the O(5) symmetry,
for instance in those described by the projected SO(5) model discussed in 
Refs.~\cite{ZHAHA-99,AH-00,DAJHZ-02}.
Of course, when the effective Hamiltonian parameters
are outside the attraction domain of the stable
FP, the transition at the MCP is expected to be of first-order type.
Some of the results concerning 
the stability properties of the O($N$) FP were
already presented in Ref.~\cite{CPV-02-so5}.

The paper is organized as follows. 
In Sec.~\ref{seceps} we present our five-loop calculations
in the framework of the $\epsilon$ expansion.
In Sec.~\ref{sec3} we discuss the stability of 
the O($N$)-symmetric
FP under generic perturbations.
The results are then applied to establish the stability properties of
the O($N$) FP.
In Sec.~\ref{sec4} the multicritical RG flow is analyzed.
In Sec.~\ref{sec5} we draw our conclusions and 
discuss their relevance for some physical systems.

\section{$\epsilon$ expansion of 
the O($\lowercase{n}_1$)$\oplus$O($\lowercase{n}_2$) theory}
\label{seceps}

We extended the $\epsilon$ expansion of the critical exponents at the 
different FP's for the O($n_1$)$\oplus$O($n_2$) 
symmetric theory to $O(\epsilon^5)$.
For this purpose, we considered the minimal
subtraction (MS) renormalization scheme \cite{tHV-72}.
We computed the divergent part of
the irreducible two-point functions of the fields $\phi_1$ and $\phi_2$,
of the two-point correlation functions with insertions of the 
quadratic operators  $\phi_1^2$ and $\phi_2^2$, and 
of the three independent four-point correlation 
functions
$\langle \phi_1\cdot \phi_1 \; \phi_1\cdot \phi_1  \rangle$,
$\langle \phi_1\cdot \phi_1 \; \phi_2\cdot \phi_2  \rangle$, and
$\langle \phi_2\cdot \phi_2 \; \phi_2\cdot \phi_2  \rangle$.
The diagrams contributing to this calculation are a few hundreds.
We handled them with a symbolic manipulation program, which  
generated the diagrams and computed the symmetry and group factors of 
each of them. We used the results of  Ref.~\cite{KS-01}, 
where the primitive divergent parts of all integrals appearing in our 
computation are reported.
We determined the
renormalization constants $Z_{\phi_1}$ and $Z_{\phi_2}$
associated with the fields $\phi_1$ and $\phi_2$ respectively, 
the $3\times 3$ renormalization matrix $Z^g_{ij}$
of the quartic couplings defined by $g_{B,i} = \mu^\epsilon Z^g_{ij} g_{R,j}$ 
where $g_{B,i}\equiv (u_1,u_2,w)$, and the $2\times 2$ renormalization matrix
$Z^{\phi^2}_{ij}$ of the quadratic operators $\phi_1^2$ and $\phi_2^2$.
The $\beta$ functions $\beta_i(g_{R,j})$ and the RG dimensions
$\gamma_{\phi_1}$, $\gamma_{\phi_2}$, $\gamma^{\phi^2}_{ij}$  
are determined using the relations
\begin{eqnarray}
&&\beta_i(g_{R,i}) = 
   \mu \left. {\partial g_{R,i} \over \partial \mu}\right|_{g_{B,j}},\\
&&\gamma_{\phi_i}(g_{R,i}) = 
   \sum_j \beta_j {\partial Z_{\phi_i}\over \partial g_{R,j}},\\
&&\gamma^{\phi^2}_{ij} (g_{R,i}) = 
  \sum_{kl} \beta_k {\partial Z^{\phi^2}_{kl}\over \partial g_{R,i}} 
      (Z^{\phi^2})^{-1}_{lj}.
\end{eqnarray}
The zeroes $g^*_{R,i}$ of the $\beta$-functions provide the FP's
of the theory. In the framework of the $\epsilon$ expansion, they are obtained 
as perturbative expansions 
in $\epsilon$ and are then inserted in the RG functions to determine
the $\epsilon$ expansion of the critical exponents.
The stability of each FP
is controlled by the $3\times 3$ matrix 
\begin{equation}
\Omega_{ij} = 
\left. {\partial \beta_i(g_{R,k})\over \partial g_{R,j} }
        \right|_{g_{R,k}=g^*_{R,k}}.
\label{omega}
\end{equation}
The two exponents $\eta_1$ and $\eta_2$, related to
the short-distance behavior of the two-point functions 
of the fields $\phi_1$ and $\phi_2$, are given by
$\eta_1=\gamma_{\phi_1}(g^*_{R,i})$ and $\eta_2=\gamma_{\phi_2}(g^*_{R,i})$.
From the eigenvalues $\nu_1$ and $\nu_2$ of 
the matrix $\gamma^{\phi^2}_{ij}$, if $\nu_1 > \nu_2$, one obtains
$\nu=\nu_1$ and $\phi=\nu_1/\nu_2$, where $\phi$ is the crossover exponent
associated with the quadratic instability.

We performed several checks of the perturbative series. In particular,
the critical-exponent series agree with 
the existing $O(\epsilon^5)$ ones 
for the O($N$)-symmetric theory \cite{CGLT-83,KNSCL-93} in the proper limit. 
Morover, as we shall discuss in the following section, we can also compare 
with some results for the O($N$) theory 
in the presence of cubic anisotropy \cite{KS-95}, finding  agreement.
Some of the five-loop perturbative series will be reported in the following 
sections.  The complete list of series is available on request.

Since the $\epsilon$ expansion is asymptotic, the series must be properly
resummed to provide results for three-dimensional systems.
We used the Pad\'e-Borel method 
except for the series at the O($N$) FP. In this case,
we applied the conformal-mapping method \cite{LZ-77} 
that takes into account
the known large-order behavior of the expansion.
See, e.g., Refs.~\cite{Zinn-Justin-book,PV-r} for 
reviews of resummation methods.

\section{Stability of the O($N$) fixed point}
\label{sec3}

In this section we discuss the stability of the O($N$) FP,
where $N=n_1+n_2$,
to establish in which cases the enlargement of the symmetry
O($n_1$)$\oplus$O($n_2$) to  O($N$) is realized at the MCP
without the need of further tunings.

Let us consider the general problem of an
O($N$)-symmetric Hamiltonian in the
presence of a perturbation $P$, i.e.,
\begin{equation}
{\cal H} = \int d^d x \left[
\case{1}{2} ( \partial_\mu \Phi)^2  + \case{1}{2} r \Phi^2  
+ \case{1}{4!} u (\Phi^2)^2 + h_p P \right],
\end{equation}
where $\Phi$ is an $N$-component field and
$h_p$ an external field coupled to $P$.
Assuming $P$ to be an eigenoperator of the RG transformations,
the singular part of the Gibbs free energy for the reduced temperature
$t\to 0$ and  $h_p \to 0$ can be written as
\begin{equation}
{\cal F}_{\rm sing}(t,h_p) \approx |t|^{d\nu}
 \widehat{\cal F} \left(h_p |t|^{-\phi_p}\right),
\end{equation}
where $\phi_p \equiv y_p \nu$ is the crossover exponent associated with the 
perturbation $P$, $y_p$ is the RG dimension of $P$,
and $\widehat{\cal F} (x)$ is a scaling function.
If $y_p>0$ the pertubation is relevant and its presence
causes a crossover to another critical behavior or to a first-order transition.

In order to discuss the stability of the O($N$) FP  in general, 
we must consider any perturbation
of the O($N$) FP. We shall first consider perturbations that 
are polynomials of the field $\Phi^a$. Any such perturbation can be 
written \cite{Wegner-72} as a sum of terms $P_{m,\l }^{a_1,\ldots,a_{\l} }$, 
$m\ge \l$, which are 
homogeneous in $\Phi^a$ of degree $m$ and transform as the $\l$-spin 
representation of the O($N$) group. Explicitly, we have 
\begin{equation}
    P_{m,{\l}}^{a_1,\ldots,a_{\l}} = (\Phi^2)^{m-\l} Q_{\l}^{a_1,\ldots,a_{\l}}
\end{equation}
where $Q_{\l}^{a_1,\ldots,a_{\l}}$ is a homogeneous polynomial of degree $\l$ 
that is symmetric and traceless in the $\l$ indices. The lowest-order 
even polynomials are
\begin{eqnarray}
&&Q^{ab}_{2} = \Phi^a \Phi^b - {1\over N} \delta^{ab} \Phi^2
 \label{spin2}\\
&&Q^{abcd}_{4} = \Phi^a \Phi^b \Phi^c \Phi^d  
- \case{1}{N+4} \Phi^2 \left( 
        \delta^{ab} \Phi^c \Phi^d + \delta^{ac} \Phi^b \Phi^d + 
        \delta^{ad} \Phi^b \Phi^c + \delta^{bc} \Phi^a \Phi^d + 
        \delta^{bd} \Phi^a \Phi^c + \delta^{cd} \Phi^a \Phi^b \right) 
\nonumber \\
   && \qquad + \case{1}{(N+2)(N+4)} (\Phi^2)^2 \left(
         \delta^{ab} \delta^{cd} + \delta^{ac} \delta^{bd} + 
         \delta^{ad} \delta^{bc} \right).
\label{spin4} 
\end{eqnarray}
The classification in terms of spin values is particularly convenient,
since polynomials with different spin do not mix under RG transformations. 
On the other hand, operators with different $m$ but with the same $\l$ do mix 
under renormalization. At least near four dimensions, we can use standard 
power counting to verify that the perturbation with indices $m,\l$ mixes only 
with $P_{m',l}$, $m'\le m$. In particular, $P_{l,l}$ renormalizes
multiplicatively and is therefore a RG eigenoperator. Moreover, if 
$y_{m,\l}$ is the RG dimension of the appropriately subtracted $P_{m,\l}$,
one can verify that for small $\epsilon$, $y_{m,\l} < 0$, for $l \ge 5$, i.e.
the only relevant operators have $\l \le 4$. We will assume this property
to hold up to $\epsilon=1$. We notice that it is certainly incorrect 
in two dimensions where perturbations are relevant ($N\ge 3$) or marginal
($N=2$) for all values of $\l$.\cite{footnoteON-2d} 
In principle, we should also consider terms with derivatives of the field, 
but again, using power counting, one can show that they are all irrelevant 
or redundant. Therefore, beside the O($N$)-symmetric terms $\Phi^2$ and 
$(\Phi^2)^2$ there are only three other perturbations that must be 
considered, $P_{2,2}^{ab}$, $P_{4,2}^{ab}$, and $P_{4,4}^{abcd}$. 
Note that, according to the above-reported discussion, $P_{2,2}^{ab}$ and 
$P_{4,4}^{abcd}$ are RG eigenoperators, while $P_{4,2}^{ab}$ 
must be in general properly subtracted, i.e. the 
RG eigenoperator is $P_{4,2}^{ab} + z P_{2,2}^{ab}$ for a suitable value of $z$.
The determination of the mixing coefficient $z$ represents a subtle point in 
the fixed-dimension expansion \cite{Parisi-80}, but is trivial
in the MS scheme in $4-\epsilon$ dimensions, in which
operators with different dimensions never mix so that $z=0$. 

According to the above-presented general analysis, the stability properties 
of the O($N$) FP can be obtained by determining the RG dimensions of the 
five operators reported above. Of course, the result does not depend on the 
specific values of the indices and thus one can consider any particular 
combination.  We now show that such dimensions determine 
the crossover exponent $\phi$ and the eigenvalues of the stability 
matrix $\Omega$ at the O($N$) FP  for the O($n_1$)$\oplus$O($n_2$) theory. 
Starting from the general expressions, one can construct combinations 
that are invariant under the symmetry group O($n_1$)$\oplus$O($n_2$). 
Explicitly, they are given by
\begin{eqnarray}
&&{\cal P}_{2,0}= \Phi^2,\qquad  \qquad  
{\cal P}_{2,2}= \sum_{a=1}^{n_1} P_{2,2}^{aa} = 
                \phi_1^2-{n_1\over N} \Phi^2 , \nonumber \\
&&{\cal P}_{4,0}= (\Phi^2)^2,\qquad  \qquad  
{\cal P}_{4,2}= \Phi^2 {\cal P}_{2,2},\nonumber \\
&&{\cal P}_{4,4}=  \sum_{a=1}^{n_1} \sum_{b=n_1+1}^{n_2} P_{4,4}^{aabb} = 
\phi_1^2 \phi_2^2- {\Phi^2 (n_1 \phi_2^2+n_2 \phi_1^2)\over N+4}+
{n_1 n_2 (\Phi^2)^2 \over (N+2)(N+4)}.
\end{eqnarray}
Here $\Phi$ is the $N$-component field $(\phi_1,\phi_2)$.
The RG dimensions of ${\cal P}_{2,0}$ and of ${\cal P}_{4,0}$ are well-known 
and can be computed directly in the O($N$)-invariant theory. In
particular, $y_{2,0}=1/\nu$ and $y_{4,0}=-\omega$, where $\omega$ 
is the leading irrelevant exponent in the O($N$)-invariant theory.
The RG dimension $y_{2,2}$ of ${\cal P}_{2,2}$, and therefore of the 
operator $P_{2,2}^{ab}$, provides the crossover exponent $\phi=y_{2,2}\nu$
at the MCP.  We denote such exponent by $\phi_T$ to stress the fact that 
it is associated with the {\em tensor} quadratic operator.
Setting
\begin{equation}
\phi_T = 1 + \sum_{i=1} p_i \epsilon^i,
\end{equation}
we obtain at five loops
\begin{eqnarray}
&&p_1 = \case{N}{2(N+8)}, \qquad\qquad 
p_2=\case{N(N^2+24N+68)}{4 (N+8)^3},  
\nonumber \\
&&p_3 = \case{N(N^4 +48N^3+788N^2+3472N + 5024)}{8 (N+8)^5}- 
       \case{6N(5N+22)\zeta(3)}{(N+8)^4},   \nonumber \\
&&p_4 = 
\case{N(N^6+72 N^5+2085N^4+28412N^3+147108N^2+337152 N + 306240)}{16(N+8)^7}
+ \case{N(-N^4+13 N^3-544N^2-4716N-8360)\zeta(3)}{(N+8)^6}\nonumber\\
&&\quad - \case{N(5N+22)\pi^4}{20(N+8)^4} 
+\case{20N(2N^2+ 55 N + 186)\zeta(5)}{(N+8)^5}
\nonumber \\
&&p_5=
\case{N\,( 17677824 + 28388096\,N + 19390624\,N^2 + 6723904\,N^3 + 
       1177480\,N^4 + 95668\,N^5 + 4154\,N^6 + 96\,N^7 + N^8)}{32\,(8 + N)^9}
\nonumber\\
&& \quad
- \case{N\,(8360+4716\,N+544\,N^2-13\,N^3+N^4)\,\pi^4 }{120\,(8+N)^6}
+ \case{5\,N\,( 186 + 55\,N + 2\,N^2)\,\pi^6}{189\,(8+N)^5}
\nonumber\\
&& \quad - \case{N\,( 554064 + 465592\,N + 125232\,N^2 + 7584\,N^3 - 
         661\,N^4 + 9\,N^5)\, \zeta(3)}{(8+N)^8}  
\nonumber\\
&& \quad + \case{2\,N\,(24528 + 14468\,N + 2028\,N^2 + 39\,N^3 + 4\,N^4)\, \zeta(3)^2}
   {( 8 + N)^7}
\nonumber\\
&&\quad + \case{N\,(466016+280596\,N+33832\,N^2-2857\,N^3-230\,N^4) \,\zeta(5)}{2\,(8+N)^7}
- \case{441\,N\,( 526 + 189\,N + 14\,N^2)\, \zeta(7)}{2\,(8+N)^6}\; .
\label{phitexp} 
\end{eqnarray}
This series extends the three-loop results of Ref. \cite{Yamazaki-74} and 
the four-loop results of Ref. \cite{Kirkham-81}.
In the appropriate limit, 
it is in  agreement with the 
$O(N^{-2})$ expression of Ref. \cite{Gracey-02}. 
In Table~\ref{ONstab} we report the estimates of $y_{2,2}$
and $\phi_T$ for $N=2,3,4,5$ obtained from the analysis of the five-loop
perturbative expansion (\ref{phitexp}).
As expected, since $y_{2,2}>0$ in all cases, the quadratic 
perturbation $P_{2,2}^{ab}$ is always relevant.
The results are compared with the estimates obtained 
from the analysis of its six-loop fixed-dimension expansion \cite{CPV-02} 
and of its large-$N$ expansion to $O(1/N^2)$ \cite{Gracey-02}, and
by using high-temperature techniques \cite{PJF-74} and
Monte Carlo simulations \cite{Hu-01}.
We also mention that consistent results were obtained from the analysis of the 
four-loop series of $\phi_T$ \cite{Kirkham-81}:
$\phi_T=1.177$ for $N=2$ and $\phi_T=1.259$ for $N=3$.
Some experimental results for $\phi_T$ can be found in Ref.~\cite{PV-r}.

\begin{table}[tbp]
\caption{
Estimates of the RG dimensions $y_{2,2}$, $y_{4,0}$, $y_{4,2}$, and $y_{4,4}$,
and of the crossover exponents $\phi_T\equiv y_{2,2}\nu$, $\phi_{4,4}\equiv y_{4,4}\nu$, 
as obtained by various approaches:
$\epsilon$ expansion ($\epsilon$ exp), fixed-dimension expansion ($d=3$ exp),
high-temperature expansion (HT exp), Monte Carlo simulations (MC), 
and $1/N$ expansion ($1/N$ exp). 
Their values in the large-$N$ limit, 
see, e.g., Ref.~\protect\cite{Ma-74}, are also reported. 
}
\label{ONstab}
\begin{tabular}{llllllll}
\multicolumn{1}{c}{$N$}&
\multicolumn{1}{c}{method}&
\multicolumn{1}{c}{$y_{2,2}$}&
\multicolumn{1}{c}{$\phi_T$}&
\multicolumn{1}{c}{$y_{4,0}$}&
\multicolumn{1}{c}{$y_{4,2}$}&
\multicolumn{1}{c}{$y_{4,4}$}&
\multicolumn{1}{c}{$\phi_{4,4}$}\\
\tableline \hline
2 &$\epsilon$ exp& 1.766(6) & 1.174(12) & $-$0.802(18) \cite{GZ-98} & $-$0.624(10) &$-$0.114(4) \cite{CPV-00} &$-$0.077(3)\\
&$d=3$ exp & & 1.184(12) \cite{CPV-02} & $-$0.789(11) \cite{GZ-98} & & $-$0.103(8) \cite{CPV-00} & $-$0.069(5) \\ 
& HT exp  & & 1.175(15) \cite{PJF-74} & & &  & \\ 
& MC & & & $-$0.795(9) \cite{CHPRV-01} & &  $-$0.17(2) \cite{CH-98} & \\ \hline

3&$\epsilon$ exp& 1.790(3) & 1.260(11)& $-$0.794(18) \cite{GZ-98}& $-$0.550(14) & 0.003(4) \cite{CPV-00} & 0.002(3) \\ 
&$d=3$ exp & & 1.27(2) \cite{CPV-02} & $-$0.782(13) \cite{GZ-98} & &  0.013(6)\cite{CPV-00}& 0.009(4) \\ 
  & HT exp & & 1.250(15) \cite{PJF-74} & & & &  \\ 
& MC & & & $-$0.773 \cite{Hasenbusch-01} & &  $-$0.0007(29) \cite{CH-98} & \\ 
  &$1/N$ exp & & 1.187 \cite{Gracey-02} & & & &  \\ \hline 

4&$\epsilon$ exp& 1.813(6) & 1.329(16)& $-$0.795(30) \cite{GZ-98}& $-$0.493(14) & 0.105(6)\cite{CPV-00}& 0.079(5) \\ 
&$d=3$ exp & & 1.35(4) \cite{CPV-02} & $-$0.774(20) \cite{GZ-98} & & 0.111(4)\cite{CPV-00}& 0.083(3)  \\ 
& MC & & & $-$0.765 \cite{Hasenbusch-01} & &  0.130(24) \cite{CH-98} & \\ 
  &$1/N$ exp & & 1.323 \cite{Gracey-02} & & & &  \\ \hline

5  &$\epsilon$ exp& 1.832(8) & 1.40(3) & $-$0.783(26) & $-$0.441(13) & 0.198(11) &  0.151(9) \\ 
&$d=3$ exp & & 1.40(4) \cite{CPV-02} & $-$0.790(15) & &  0.189(10)\cite{CPV-02-so5}& 0.144(8)  \\ 
  &MC & & 1.387(30) \cite{Hu-01} & & & & \\ 
  &$1/N$ exp & & 1.422 \cite{Gracey-02} & & & & \\ \hline

$\infty$ & & 2 & 2 & $-$1 & 0 & 1 & 1 \\
\end{tabular}
\end{table}

The perturbative expansions of the RG dimensions of the operators
${\cal P}_{4,\l}$, and therefore of the more general operators $P_{4,\l}$,
can be obtained from the eigenvalues of the stability matrix
$\Omega$ at the O($N$) FP. 
For this purpose, it is convenient to perform a change of variables, 
replacing $u_1$, $u_2$, and $w$ with  $g_{\l}$, $\l =0,2,4$, which 
are the quartic couplings associated with the operators ${\cal P}_{m,\l}$
and are explicitly defined by the relation
\begin{equation}
u_1 (\phi_1^2)^2 + u_2 (\phi_2^2)^2 + 2 w \phi_1^2\phi_2^2 
= g_0 {\cal P}_{4,0}  + g_2 {\cal P}_{4,2}  + g_4 {\cal P}_{4,4}.
\end{equation} 
In this basis $\Omega$ is diagonal and 
the eigenvalues of $\Omega$ are simply given by
\begin{equation}
\omega_{\l} = 
\left. {\partial {\bar\beta}_{\l} (g_0,g_2,g_4) \over \partial g_{\l}}
\right|_{g_0=g^*_N,\;g_2=0,\;g_4=0},
\label{omegal}
\end{equation}
where $\l=0,2,4$,
${\bar\beta}_{\l}$ are the $\beta$-functions associated with the
couplings $g_{\l}$, and
$g^*_N$ is the FP value of the quartic coupling in the 
O($N$)-symmetric theory.
The critical exponent
$\omega_0$ is the leading irrelevant operator in 
the O($N$)-symmetric theory.
Its $O(\epsilon^5)$ expansion can be found in Refs.~\cite{CGLT-83,KNSCL-93};
several estimates are reported in Refs.~\cite{GZ-98,PV-r}.
The RG dimension $y_{4,\l}$ of the perturbation $P_{4,\l}$ is 
given by
\begin{equation}
y_{4,\l} = - \omega_{\l}.
\end{equation}
We report here the five-loop $\epsilon$ expansion of $y_{4,2}$ and $y_{4,4}$.
Setting 
\begin{equation}
y_{4,\l} = \sum_{i=1}  c_{\l,i} \epsilon^i, 
\label{y4ls}
\end{equation}
we have
\begin{eqnarray}
&&c_{2,1} =  - \case{8}{N+8}, \qquad\qquad 
c_{2,2} = \case{336 + 68\,N + 7\,N^2}{(8 + N)^3},
\nonumber \\
&& c_{2,3} = 
- \case{76544 + 26176\,N + 3264\,N^2 + 28\,N^3 - N^4}{4\,(8 + N)^5}
- \case{12\,( 352 + 82\,N + 7\,N^2) \,\zeta(3)}{(8 + N)^4} 
\nonumber \\
&&c_{2,4} = 
\case{( 20796416 + 10251520\,N + 2207744\,N^2 + 271328\,N^3 + 
       24824\,N^4 + 820\,N^5 + 5\,N^6) }{16\,( 8 + N )^7}
- \case{(352 + 82\,N + 7\,N^2) \,\pi^4}{10\,(8+N)^4}
\nonumber\\
&&\quad
- \case{2\,(-92928-34776\,N- 7544\,N^2-928\,N^3-67\,N^4+N^5) \,\zeta(3)}{(8+N)^6}
+ \case{80\,( 2232 + 632\,N + 60\,N^2 + N^3)\, \zeta(5)}{(8 + N)^5}
\nonumber \\
&&c_{2,5} = 
- \case{6019366912 + 3720851456\,N + 994704384\,N^2 + 135243264\,N^3 + 
6891584\,N^4 - 590816\,N^5 - 60520\,N^6 - 1732\,N^7 - 13\,N^8}{64\,(8+N)^9}
\nonumber\\
&&\quad
+ \case{(92928+34776\,N+7544\,N^2 + 928\,N^3 +67\,N^4 -N^5) \,\pi^4 }{60\,(8+N)^6}
+ \case{20\,( 2232 + 632\,N + 60\,N^2 + N^3)\, \pi^6}{189\,(8+N)^5}
\nonumber\\
&&\quad
-\case{(117872640+62925184\,N +14334912\,N^2+1577392\,N^3+67848\,N^4 - 
     1872\,N^5 + 200\,N^6 - 7\,N^7)\, \zeta(3)}{4\,( 8 + N)^8}
\nonumber\\
&& \quad
+ \case{8\,( 104832 + 100312\,N + 24994\,N^2 + 2571\,N^3 + 83\,N^4 + 
       2\,N^5 )\, \zeta(3)^2}{( 8 + N)^7}
\nonumber\\
&&\quad
- \case{(7263744+3733728\,N + 1095516\,N^2 + 170284\,N^3 + 14035\,N^4 + 
322\,N^5)\, \zeta(5)}{(8+N)^7}
- \case{ 441\,(16832 + 5590\,N + 631\,N^2 + 23\,N^3)\, \zeta(7)}{(8+N)^6}
\end{eqnarray}
and
\begin{eqnarray}
&&c_{4,1} = \case{N-4}{8 + N},\qquad\qquad
c_{4,2}=\case{152 + 14\,N + 5\,N^2}{(8 + N)^3}, 
\nonumber \\
&&c_{4,3} = 
- \case{17024 - 1568\,N - 1464\,N^2 - 398\,N^3 - 13\,N^4}{4\,(8+N)^5}
- \case{48\,( 46 + 7\,N + N^2) \,\zeta(3)}{(8+N)^4},
\nonumber \\
&& c_{4,4} = 
- \case{( 2995712 + 402304\,N + 223328\,N^2 + 112856\,N^3 + 
       27272\,N^4 + 1516\,N^5 + 29\,N^6)}{16\,(8+N)^7}
- \case{2\,(46 +7\,N +N^2)\, \pi^4 }{5\,(8+N)^4}
\nonumber\\
&&\quad
- \case{3\,(-21568+1664\,N+1592\,N^2+256\,N^3-8\,N^4 + N^5)\, \zeta(3) }{(8+N)^6}
+ \case{120\,( 712 + 130\,N + 13\,N^2) \,\zeta(5) }{(8+N)^5},
\nonumber\\
&& c_{4,5}= 
\case{-365813760 + 95377408\,N + 75546624\,N^2 + 35042816\,N^3 + 
11477472\,N^4 + 2184488\,N^5 + 148600\,N^6 + 4712\,N^7 + 61\,N^8
       }{64\,(8+N)^9}
\nonumber\\
&&\quad
- \case{(-21568 + 1664\,N + 1592\,N^2 + 256\,N^3 - 8\,N^4 + N^5)\, \pi^4}
   {40\,(8 + N)^6}
+ \case{10\,( 712 + 130\,N + 13\,N^2) \,\pi^6 }{63\,( 8 + N )^5}
\nonumber\\
&& \quad
- \case{(37827072+ 13773568\,N + 3633344\,N^2+689728\,N^3+54184\,N^4 - 
     3272\,N^5 + 188\,N^6 - 5\,N^7) \,\zeta(3)}{4\,(8+N)^8}
\nonumber\\
&&\quad
+ \case{12\,(11456+24112\,N+6648\,N^2+790\,N^3+5\,N^4) \,\zeta(3)^2}{(8+N)^7}
\nonumber\\
&&\quad
- \case{2\,( 1018944 + 128152\,N - 3060\,N^2 - 9018\,N^3 +347\,N^4 + 
       12\,N^5) \,\zeta(5) }{(8+N)^7}
- \case{2646\,( 1268 + 272\,N + 25\,N^2 + N^3) \, \zeta(7)}{(8+N)^6}\; .
\label{c4i} 
\end{eqnarray}
At one loop, these results agree with those reported in Ref. \cite{Wegner-72}.
The results of the analyses of these series are reported in Table~\ref{ONstab}.
They show that $y_{4,2}$ is always negative, so that the corresponding
spin-2 perturbation $P_{4,2}^{ab}$ is always irrelevant.
On the other hand, the sign of $y_{4,4}$ depends on $N$:
it is clearly negative for $N=2$ and positive for $N\ge 4$.
For $N=3$ it is marginally positive, suggesting the instability
of the O(3) FP. This fact will be 
confirmed by the more accurate results discussed below.
The corresponding crossover exponents $\phi_{m,\l}\equiv y_{m,\l}\,\nu$
can be determined using the following estimates of 
$\nu$:
$\nu = 0.67155(27)$ for $N=2$ \cite{CHPRV-01},
$\nu = 0.7112(5)$ for $N=3$ \cite{CHPRV-02},
$\nu = 0.749(2)$ for $N=4$ \cite{Hasenbusch-01},
and $\nu = 0.762(7)$ for $N=5$ \cite{O5exp}.
Other estimates of $\nu$ can be found in Ref.~\cite{PV-r}.

The RG dimension $y_{4,4}$ can also be obtained
starting from the cubic-symmetric LGW Hamiltonian, 
\begin{equation}
{\cal H}_{c} =  \int d^d x\, \left\{ {1\over 2} \sum_{i=1}^{N}
      \Bigl[ (\partial_\mu \Phi_i)^2 +  r \Phi_i^2 \Bigr]  
+{1\over 4!} \Bigl[ u (\sum_i^N \Phi_i^2)^2 + v \sum_i^N \Phi_i^4 \Bigr]
\right \},
\label{Hphi4cubic}
\end{equation}
see, e.g., Ref.~\cite{Aharony-76}, and in particular from the results for 
the stability properties of the O($N$) FP in the presence of 
a cubic-symmetric anisotropy.  The point is that the cubic-symmetric 
perturbation is
a particular combination of the spin-4 operators $P_{4,4}^{abcd}$ and 
of the spin-0 term $(\Phi^2)^2$.  Indeed, one may rewrite 
\begin{equation}
\sum_{i=1}^N\Phi^4_i = \sum_{a=1}^N P^{aaaa}_{4,4} + 
   {3 \over N+2} (\Phi^2)^2.
\end{equation}
Thus, the stability of the O($N$) FP against the cubic-symmetric perturbation
$\sum_i \Phi_i^4$ is controlled by the RG dimension $y_{4,4}$
of the spin-4 operator $P_{4,4}^{abcd}$.

The RG flow for the cubic-symmetric theory 
has been investigated by employing
field-theoretical methods, based on perturbative expansions 
\cite{KS-95,CPV-00,CPV-02-2,FHY-00,MSS-89,Varnashev-00,KTS-97,SAS-97}
or approximate solutions of continuous RG equations \cite{NR-82,TMVD-01,YH-77}, 
and lattice techniques,
such as Monte Carlo simulations \cite{CH-98}  
and high-temperature expansions \cite{FVC-81};
see, e.g., Ref.~\cite{PV-r} for a recent review.
In particular, 
the RG functions have been computed
to five loops in the $\epsilon$ expansion \cite{KS-95},
and to six loops in a fixed-dimension expansion in powers of 
the zero-momentum quartic couplings \cite{CPV-00}.
In these perturbative schemes
\begin{equation}
  y_{4,4} = - 
\left. {\partial \beta_v (u,v) \over \partial v}\right|_{u=g^*_N,\;v=0}\; ,
\label{y44b}
\end{equation}
where $\beta_v$ is the $\beta$-function associated with 
the quartic coupling $v$,
and $g^*_N$ is the FP value of the quartic coupling
in the O($N$)-symmetric theory.
This allows us to determine $y_{4,4}$ using the five-loop expansions
reported in Ref.~\cite{KS-95}. We reobtain again Eq. (\ref{c4i}), 
confirming the correctness of our calculation.
Moreover, using Eq. (\ref{y44b}) and the results reported in 
Ref.~\cite{CPV-00}, one can also compute $y_{4,4}$ in the framework
of the  fixed-dimension expansion to six loops.
The resulting estimates, obtained by using the  conformal-mapping method,
are reported in Table~\ref{ONstab}.
They show that the spin-4 perturbation $P_{4,4}$ is relevant for all $N\ge 3$.
In Table~\ref{ONstab} the Monte Carlo results of Ref.~\cite{CH-98} are also
shown; they were
obtained by simulating the standard $N$-vector model and 
computing the RG dimension of  the cubic-symmetric term
$\sum_{i} s_i^4$, where $s_i$ is the $N$-component spin variable.
We may also consider 
the value $N_c$ such that for $N>  N_c$ the cubic-symmetric
anisotropy, and therefore the spin-4 perturbation $P_{4,4}^{abcd}$,
becomes relevant at the O($N$) FP. 
All studies reported in the literature indicate $N_c\approx 3$
and definitely $N_c<4$, see, e.g.,
Refs.~\cite{CPV-00,KS-95,FHY-00,MSS-89,Varnashev-00,SAS-97,%
NR-82,TMVD-01,YH-77,CH-98,FVC-81}.
The most accurate results have been provided by analyses of 
high-order perturbative field-theory expansions, which predict  
$N_c\lesssim 2.9$ in three dimensions.
In particular, different analyses of the six-loop fixed-dimension series 
yielded the estimates  $N_c=2.89(4)$ \cite{CPV-00} 
and $N_c=2.862(5)$ \cite{FHY-00}; similar results were also obtained
from shorter series, see, e.g., Refs. \cite{MSS-89,Varnashev-00}.
These results have been confirmed by the analysis
of the $O(\epsilon^5)$ series \cite{KS-95,SAS-97,CPV-00}.
A constrained analysis taking into account the
two-dimensional value of $N_c$, $N_c=2$,
provided the estimate $N_c=2.87(5)$ \cite{CPV-00},  
which makes the evidence supporting $y_{4,4}>0$ for $N=3$
stronger than the estimate $y_{4,4}=0.003(4)$ obtained from the direct analysis
of its $O(\epsilon^5)$ series.

In conclusion, these results provide a rather robust evidence
that for $N\ge 3$ the O($N$) FP is unstable with respect to
spin-4 perturbations $P_{4,4}^{abcd}$,
and, as a consequence, that 
the O($N$) FP is a unstable MCP for $N\ge 3$.

\section{RG flow at the multicritical point}
\label{sec4}

As already shown by the $O(\epsilon)$ computations of Ref.~\cite{KNF-76},
the O($n_1$)$\oplus$O($n_2$) theory at the MCP  has six FP's. 
Three of them, i.e. the Gaussian, the O($n_1$) and the O($n_2$) FP's,  
are always unstable. The other three FP's are 
the O($N$) fixed point (FP), 
the biconal fixed point (BFP), and the 
decoupled fixed point (DFP).
The stability of these FP's depends on $n_1$ and $n_2$.
In particular, in the preceding section we have established
that the O($N$) FP is stable for $N=2$ and unstable for
$N\ge 3$, for any $n_1$ and $n_2$.

The stability properties of the DFP can be determined using
nonperturbative scaling arguments 
\cite{Aharony-76,GT-83,Aharony-02,Aharony-02-2}.
At the DFP, the quartic coupling term $w\phi_1^2\phi_2^2$ scales as
the product of two energy-like operators, which have RG dimensions
$(1-\alpha_i)/\nu_i$ where $\alpha_i$ and $\nu_i$ are the critical
exponents of the O($n_i$) universality classes. Therefore, 
the RG dimension related to the $w$-perturbation is given by
\begin{equation}
y_w = {\alpha_1\over 2\nu_1} + {\alpha_2\over 2\nu_2} = 
  {1\over \nu_1} + {1 \over \nu_2} - d.
\label{yw}
\end{equation}
Note that this relation is satisfied order by order
in the $\epsilon$ expansion.
Indeed, the $\epsilon$ expansion of $y_w$ obtained from the
stability matrix $\Omega$ at the DFP coincides with 
the series obtained from the right-hand side of Eq.~(\ref{yw}),
using the five-loop expansions of $\nu_i$
for the O($n_i$) universality classes.
Taking into account that the DFP is stable with respect
to the other two RG directions, 
one can determine the stability properties of the DFP
from the sign of $y_w$. 
Using the estimates of the critical exponents of the three-dimensional
O($n_i$) universality classes (see, e.g., Ref.~\cite{PV-r} for a review),
$y_w$ turns out to be negative for $N\equiv n_1+n_2\ge 4$, 
and positive for $N=2,3$.\cite{footnote-DFP-2d}
Three-dimensional estimates of $y_w$ for $N\le 5$ are reported in 
Table~\ref{ywtab}.  These results show that the 
tetracritical DFP is stable for $N\ge 4$ for any $n_1,n_2$.

\begin{table}[tbp]
\caption{Estimates of the RG dimension $y_w$ at the DFP.
They are obtained by using 
Eq.~(\protect\ref{yw})  and 
the estimates of the O($N$) critical exponent $\nu$  reported in 
Refs.~\protect\cite{CPRV-99,CPRV-02,CHPRV-01,CHPRV-02,Hasenbusch-01}.
}
\label{ywtab}
\begin{tabular}{cccl}
\multicolumn{1}{c}{$N=n_1+n_2$}&
\multicolumn{1}{c}{$n_1$}&
\multicolumn{1}{c}{$n_2$}&
\multicolumn{1}{c}{$y_w$}\\
\tableline \hline
2 & 1 & 1 & 0.1740(8) \\
3 & 1 & 2 & 0.0761(7) \\
4 & 1 & 3 & $-$0.0069(11) \\
  & 2 & 2 & $-$0.0218(12) \\
5 & 1 & 4 & $-$0.078(4) \\
  & 2 & 3 & $-$0.1048(12) \\
\end{tabular}
\end{table}

The results concerning the O($N$) FP and the DFP 
suggest that the stable FP for $N=3$ is the BFP.
This is substantially confirmed by the five-loop analysis of
the stability matrix $\Omega$ at the BFP.
Below we report the expansions of the critical exponents at the BFP 
for $n_1=1$ and $n_2=2$.
The eigenvalues of the stability matrix $\Omega$ are
\begin{eqnarray}
&&\omega_{{\rm bi},1} = 
\,\epsilon - 0.579364 \,\epsilon^2  + 1.344815 \,\epsilon^3  
- 4.058162 \,\epsilon^4  + 14.526420 \,\epsilon^5 + O(\epsilon^6),
\nonumber \\
&&\omega_{{\rm bi},2} =
0.491105 \,\epsilon  - 0.084149 \,\epsilon^2  + 0.361174 \,\epsilon^3  
- 0.776741 \,\epsilon^4 + 2.593212 \,\epsilon^5 + O(\epsilon^6),
\nonumber \\
&&\omega_{{\rm bi},3}=
-0.130195 \,\epsilon + 0.278782 \,\epsilon^2  - 0.379711 \,\epsilon^3   
+ 0.868886 \,\epsilon^4  - 2.656984 \,\epsilon^5 + O(\epsilon^6).
\end{eqnarray}
The expansions of the critical exponents are 
\begin{eqnarray}
&&\eta_{{\rm bi},1} = 
0.0208306 \,\epsilon^2  + 0.0182941 \,\epsilon^3  - 0.00777325 \,\epsilon^4  
+ 0.0210296 \,\epsilon^5  + O(\epsilon^6),
\nonumber \\
&&\eta_{{\rm bi},2} = 
0.0200806 \,\epsilon^2  + 0.0195184 \,\epsilon^3  
- 0.00848020 \,\epsilon^3  + 0.0236398 \,\epsilon^5  + O(\epsilon^6),
\nonumber \\
&&\nu_{{\rm bi},1} = 
\case{1}{2} + 0.1114875 \,\epsilon + 0.0667684 \,\epsilon^2  
- 0.00616190 \,\epsilon^3  + 0.0779498 \,\epsilon^4  - 0.193367 \,\epsilon^5
+ O(\epsilon^6),
\nonumber \\
&&\nu_{{\rm bi},2} =
\case{1}{2} + 0.0234143 \,\epsilon + 0.0289670 \,\epsilon^2  - 0.00547548 \,\epsilon^3  
+ 0.0381483 \,\epsilon^4  - 0.106076 \,\epsilon^5 + O(\epsilon^6),
\nonumber \\
&&\phi_{\rm bi} = {\nu_{{\rm bi},1}\over \nu_{{\rm bi},2}} = 
1 + 0.176147 \,\epsilon + 0.0673541 \,\epsilon^2  - 0.0147318 \,\epsilon^3  
\nonumber \\[-3mm]
&&\hphantom{\phi_{\rm bi} = {\nu_{{\rm bi},1}\over \nu_{{\rm bi},2}} =}
+ 0.0783198 \,\epsilon^4  - 0.190099 \,\epsilon^5 + O(\epsilon^6).
\end{eqnarray} 
We analyzed these series 
using the Pad\'e-Borel resummation method.
The estimates of the eigenvalues of the stability matrix are
$\omega_{{\rm bi},1}=0.79(2)$, $\omega_{{\rm bi},2}=0.57(4)$, and 
$\omega_{{\rm bi},3}=0.01(1)$.
They are all positive, supporting the stability of the BFP,
although the result for $\omega_{{\rm bi},3}$ is not sufficiently precise
to definitely exclude the opposite sign.
Concerning the critical exponents, we obtained
$\eta_{{\rm bi},1}=0.037(5)$, 
$\eta_{{\rm bi},2}=0.037(5)$, 
$\nu_{\rm bi}=\nu_{{\rm bi},1}=0.70(3)$, and
$\phi_{\rm bi}=1.25(1)$.
Note first that $\eta_{{\rm bi},1}\approx \eta_{{\rm bi},2}$, as it can be 
directly guessed by looking at the coefficients of their expansions. 
A direct analysis of their difference gives the bound 
$|\eta_{{\rm bi},1}- \eta_{{\rm bi},2}|\lesssim 0.0005$.
Second, note that, within the errors, the BFP exponents are very close to 
the Heisenberg ones, whose best estimates are
$\eta_{H}=0.0375(5)$, $\nu_{H}=0.7112(5)$, and $\phi_{H}=1.250(15)$ 
from high-temperature techniques \cite{CHPRV-02,PJF-74}, 
$\eta_{H}=0.0355(25)$, $\nu_{H}=0.7073(35)$, and $\phi_{H}=1.27(2)$ 
from the six-loop fixed-dimension expansion \cite{GZ-98,CPV-02},
$\eta_{H}=0.0375(45)$, $\nu_{H}=0.7045(55)$, and $\phi_{H}=1.260(11)$
from the five-loop $\epsilon$ expansion \cite{GZ-98}.
Rather stringent bounds on the differences 
between the biconal and Heisenberg exponents can be obtained by
considering the expansions of their differences, which have 
much smaller coefficients. Their analysis yields
\begin{eqnarray}
&&  |\eta_{{\rm bi},1}-  \eta_{H}| \lesssim  0.0005, \nonumber \\
&&|\eta_{{\rm bi},2}-  \eta_{H}| \lesssim 0.0001,\nonumber \\
&&|\nu_{\rm bi} - \nu_{H}| \lesssim 0.001,\nonumber \\
&&|\phi_{\rm bi} - \phi_{H}| \lesssim 0.005. \label{diffexp}
\end{eqnarray}

We have also studied the stability of the BFP for larger values of $N$.
For $N=4$, and in both cases $n=1$, $n_2=3$ and $n_1=n_2=2$,
the five-loop calculation gives the expansions of the critical 
exponents at the BFP only
to $O(\epsilon^4)$, because of the additional degeneracy of the O(4) FP and 
of the BFP at $O(\epsilon)$. 
In particular, for the smallest eigenvalue we obtain 
\begin{eqnarray}
&&\omega_{{\rm bi},3}(n_1=1,n_2=3)=
\case{1}{6} \epsilon^2  - 0.3306439 \,\epsilon^3
+ 0.7376491 \,\epsilon^4  + O(\epsilon^5),
\nonumber \\
&&\omega_{{\rm bi},3}(n_1=2,n_2=2)=
\case{1}{6} \epsilon^2  - 0.319872 \,\epsilon^3
+ 0.696458 \,\epsilon^4  +  O(\epsilon^5).
\end{eqnarray}  
It is difficult to extract reliable estimates from these series. 
In both cases, we find that $\omega_{{\rm bi},3}$ is small, 
but we are unable to determine reliably its sign.

For $N\ge 5$ we find
that the BFP is unstable for all values of $n_1$ and $n_2$.
In particular, for $N=5$, $n_1 = 2$, $n_2 = 3$, 
for the smallest eigenvalue we obtain
\begin{eqnarray}
&&\omega_{{\rm bi},3}=
0.052584 \epsilon  + 0.0331401 \,\epsilon^2
- 0.242179 \,\epsilon^3
+ 0.358964 \,\epsilon^4  - 1.242100 \,\epsilon^5 + O(\epsilon^6),
\end{eqnarray}
which gives $\omega_{{\rm bi},3} = -0.07(5)$.

\section{Conclusions and discussion}
\label{sec5}

We have studied the multicritical behavior at a MCP,
where two critical lines with O($n_1$) and O($n_2$) symmetry meet.
It has been  determined by 
studying the RG flow of the most general O($n_1$)$\oplus$O($n_2$)-symmetric
LGW Hamiltonian involving two fields $\phi_1$ and $\phi_2$
with $n_1$ and $n_2$ components respectively. 
We have extended the $\epsilon$ expansion of the critical exponents 
and of the stability matrix of the FP's, previously known 
to one-loop order, to five loops.
The stability of the O($N$) FP has
also been discussed in the framework of the fixed-dimension expansion in
three dimensions to six loops.

The main properties of the RG flow of the O($n_1$)$\oplus$O($n_2$)-symmetric
system at the MCP can be summarized as follows.

\begin{itemize}
\item
The O($N$) FP is stable only for $N=2$, i.e.
when two Ising-like critical lines meet.
It is  unstable in all other cases, i.e. for all 
$n_1$ and $n_2$ such that $n_1 + n_2 = N\ge 3$.
Beside being unstable with respect to the spin-0 and
spin-2 quadratic perturbations, for $N\ge 3$ the O($N$) FP is also unstable 
with respect to quartic perturbations belonging to the
spin-4 representation of the O($N$) group, cf. Eq.~(\ref{spin4}).
This implies that for $N\ge 3$ the enlargement of the symmetry 
O($n_1$)$\oplus$O($n_2$) to O($N$)
requires an additional parameter to be tuned,
beside those associated with the quadratic perturbations, 
$r_1$ and $r_2$ in the LGW Hamiltonian.
The associated crossover exponents $\phi_{4,4}\equiv y_{4,4}\,\nu$ 
are:
$\phi_{4,4}\approx 0.01$ for $N=3$,
$\phi_{4,4}\approx 0.08$ for $N=4$, 
$\phi_{4,4}\approx 0.15$ for $N=5$,
and $\phi_{4,4}\rightarrow 1$ for $N\rightarrow \infty$ 
(see Table~\ref{ONstab}).

\item
For $N=3$, i.e. for $n_1=1$ and $n_2=2$,
the critical behavior at the MCP is described by the BFP,
whose critical exponents turn out to be very close to those of
the Heisenberg universality class, see Eq.~(\ref{diffexp}).

\item
For $N\ge 4$ and for any $n_1\ge 1$ and $n_2\ge 1$,
the tetracritical DFP is stable.
This has been inferred using nonperturbative arguments 
\cite{Aharony-76,GT-83,Aharony-02,Aharony-02-2}
that allow us to write the relevant stability eigenvalue
$y_w$ in terms of the critical exponents of the O($n_i$)
universality classes, cf. Eq.~(\ref{yw}).
The $\epsilon$-expansion analysis
shows that the BFP is unstable for all cases with 
$N\ge 5$, while it is not conclusive for the cases 
with $N=4$.

\item
When the initial parameters of the Hamiltonian
are not in the attraction domain
of the stable FP, the transition between the disordered
and ordered phases should be of first order in the neighborhood 
of the MCP. 
In this case, a possible phase diagram is given in Fig.~\ref{tricr}.
Close to the MCP all transition lines are first-order ones. However,
far from the MCP, the high-temperature transitions may become continuous,
belonging to the O($n_1$) and O($n_2$) universality classes.
\end{itemize}

As already mentioned in the introduction, a multicritical behavior
has been observed in several systems.  

Anisotropic antiferromagnets in a uniform magnetic field $H_\parallel$
parallel to the anisotropy axis present a MCP 
in the $T-H_\parallel$ phase diagram, where two critical lines
belonging to the XY and Ising universality classes meet \cite{NKF-74,KNF-76}. 
The results presented above predict a multicritical BFP.
The mean-field approximation assigns a tetracritical behavior to the MCP
\cite{KNF-76}, but a more rigorous characterization, that requires
the computation of the corresponding scaling free energy, is needed 
to draw a definite conclusion.
Experimentally, the MCP appears to be bicritical, see, e.g., the experimental
results of Refs. \cite{RG-77,KR-79}; numerical Monte Carlo results hint at the 
same behavior, although with much less confidence \cite{LB-78}.
Our results contradict the $O(\epsilon)$ 
calculations of Refs. \cite{NKF-74,KNF-76}, suggesting the
stability of the O(3) FP.
Notice that it is very hard to distinguish 
the biconal from the  O(3) critical behavior.
For instance, the correlation-length exponent
$\nu$ differs by less than 0.001 in the two cases.
However, one may still hope to distinguish the two  FP's
by measuring some universal amplitude ratio that varies more significantly 
in the two cases.
The crossover exponent describing the crossover from the O(3)
critical behavior 
is very small, i.e., $\phi_{4,4}\approx 0.01$,
so that systems with a small effective breaking
of the O(3) symmetry cross very slowly towards
the biconal critical behavior or,
if the system is outside the attraction domain of the BFP,
towards a first-order transition; thus, they may 
show the eventual asymptotic behavior only
for very small values of the reduced temperature.

Isotropic antiferromagnets in a magnetic field are quite a special case.
Indeed, the critical transition at $H=0$ is a MCP with O(3) symmetry, 
as observed experimentally, see, e.g., Ref. \cite{SB-77}. As we discussed, 
for $H\not=0$, two relevant perturbations are switched on, and they can give 
rise in principle to a more complex phase diagram.
Finally, it should be noted that in real antiferromagnets 
additional nonisotropic interactions are present, giving rise to lower-symmetry
MCP's. In Ref. \cite{BOPFS-88} the magnetic phase diagram of 
NiCl${}_2$$\cdot$4H$_2$O was studied. The orthorombic symmetry of the 
crystal gives rise to Ising transition lines both for small and large 
$H_\parallel$, so that $n_1=n_2=1$. As predicted by the theory, the MCP 
is a bicritical XY point. A similar experiment is reported in 
Ref. \cite{BFdJ-80}. A tetracritical XY MCP is observed in 
anisotropic antiferromagnets when the magnetic field is perpendicular
to the symmetry axis, see, e.g., Ref. \cite{RG-77} for an experimental study.

High-$T_c$ superconductors  are
other interesting physical systems
in which MCP's may arise
from the competition of different order parameters.
At low temperatures these materials exhibit 
superconductivity and antiferromagnetism
depending on doping. The SO(5) theory \cite{Zhang-97,ZHAHA-99} 
attempts to provide a unified description of these
two phenomena, involving a three-component
antiferromagnetic order parameter and 
a $d$-wave superconducting order parameter with U(1) symmetry,
with an approximate O(5) symmetry.
This theory predicts 
a MCP arising from the competition of these two 
order parameters when the corresponding critical lines meet
in the temperature-doping phase diagram.
Neglecting the fluctuations of the magnetic field
and the quenched randomness introduced by doping, see,
e.g., Ref.~\cite{Aharony-02-2} for a critical discussion of this point,
one may consider  the O(3)$\oplus$O(2)-symmetric LGW Hamiltonian 
to infer the critical behavior at the MCP, see, e.g.,
Refs. \cite{BL-98,AH-00,HZ-00,MN-00,Hu-01,KAE-01}.
In particular, the analysis of Ref.~\cite{AH-00},
which uses the projected SO(5) model \cite{footnote1}
as a starting point, 
shows that one can use Eq. (\ref{bicrHH}) as an effective 
Hamiltonian. 
Different scenarios have been proposed for the 
critical behavior at the MCP. 
In Refs.  \cite{Zhang-97,HZ-00,Hu-01}, it was speculated that the 
MCP is a bicritical point where the O(5) symmetry is asymptotically 
realized. 
On the other hand, on the basis of the $O(\epsilon)$ results 
of Refs.~\cite{NKF-74,KNF-76},
Refs.~\cite{BL-98,MN-00} 
predicted a tetracritical behavior governed by the
BFP.  However, since it was expected that the BFP
is close to the O(5) FP, it was suggested that
at the MCP the critical exponents were in any case close to 
the O(5) ones.

The O(5)-symmetric scenario would require
the stability of the O(5) FP.
Evidence in favor of this picture has been recently
claimed using Monte Carlo simulations for a five-component 
O(3)$\oplus$O(2)-symmetric spin model \cite{Hu-01,Hu-02}.
The numerical results show that, within the parameter
ranges considered, the scaling behavior at the MCP
is consistent with an O(5)-symmetric critical behavior.
Similar results have been obtained in Ref.~\cite{DAJHZ-02}
by a quantum Monte Carlo study of the quantum projected
SO(5) model in three dimensions. 
On the other hand,
the interpretation of these numerical results as an evidence
for the stability of the O(5) FP \cite{Hu-01,Hu-02}
is untenable, because 
the results discussed in this paper
definitely show that the O(5) FP is unstable, and that
the asymptotic approach to the MCP is characterized
by a decoupled critical behavior or by a first-order transition.
The O(5) symmetry can be asymptotically realized only by tuning 
a further relevant parameter, beside the double tuning  required to approach 
the MCP. We note that the crossover exponent $\phi_{4,4}$,
related to the spin-4 pertubation of the O(5) FP,
$\phi_{4,4}\approx 0.15$, is much larger than
its $O(\epsilon)$  approximation, i.e., 
$\phi_{4,4}\approx\case{1}{26} \epsilon$
from which one would obtain $\phi_{4,4}\approx 0.04$ setting $\epsilon=1$. 
It is of the same order of the crossover exponent
appearing in many other physical systems. 
For instance, in randomly dilute uniaxial magnetic materials---a class 
of systems whose asymptotic critical behavior has been precisely observed 
both numerically and experimentally, see, 
e.g, Refs. \cite{PV-r,Belanger-00} for reviews---the pure Ising fixed point is unstable with 
crossover exponent $\phi\approx0.11$, which is even smaller than the 
above-reported estimate for the O(5) case. Therefore, 
contrary to some recent claims \cite{DAJHZ-02}, it cannot be excluded 
{\em a priori} that experiments are able to observe the unstable flow out 
of the O(5) fixed point, even in those systems with a moderately
small breaking of the O(5) symmetry, such as the projected SO(5) model.
Evidence in favor of a tetracritical behavior has been recently
provided by a number of experiments, see, e.g.,
Refs.~\cite{Khay-etal-02,Lee-etal-99,Katano-etal-00,Miller-etal-02,%
Lake-etal-01,Aeppli-etal-97},
which seem to show the existence of a coexistence region 
of the antiferromagnetic and superconductivity phases.
The possible coexistence of the two phases has been discussed
in Refs.~\cite{ZDS-02,KAE-01,MOBB-00}.

Finally, we mention that a multicritical behavior with
two XY order parameters 
is expected in liquid crystals, at 
the nematic--smectic-A--smectic-C multicritical
point \cite{GT-83} and in the presence of ferromagnetic and nematic 
interactions  \cite{LGT-86}.



\begin{references}
\bibitem{CCBW-footnote}
The competing order parameters are the four-component
order parameter for chiral symmetry breaking and
the six-component order parameter for color symmetry breaking.
\bibitem{ahooo}
A different claim has been recently reported in Ref.~\cite{DAJHZ-02},
starting from the projected SO(5) model. 
Using the typical values of the  parameters characterizing 
the breaking of the O(5) symmetry and 
the crossover exponent at the O(5) FP, 
they argued that crossover effects from the O(5) 
behavior should be observable only for 
$t\equiv (T-T_c)/T_c \lesssim t_{\rm cross} \approx 10^{-10}$, i.e.,
in an extremely small interval of reduced temperatures. 
This fact led the authors to conclude
that the multicritical behavior is effectively dominated by the
O(5) FP and that it is practically impossible to observe both 
experimentally and numerically the flow out of the symmetric O(5) FP.
In order to obtain the above-reported estimate of $t_{\rm cross}$,
they computed the crossover exponent from its $O(\epsilon)$ 
approximation \cite{KNF-76}, i.e., $y_{4,4}=\case{1}{13}\epsilon + O(\epsilon^2)$
and $\phi_{4,4} = \case{1}{26}\epsilon + O(\epsilon^2)$,
thus $\phi_{4,4}\approx \case{1}{26}$ setting $\epsilon=1$.
Unfortunately, the actual value of
$\phi_{4,4}$ is much larger: $\phi_{4,4}\approx 0.15$. 
Taking for granted the arguments leading to the estimate
$t_{\rm cross} \approx 10^{-10}$, 
one can easily see that, using the actual value of 
$\phi_{4,4}$, the crossover reduced temperature $t_{\rm cross}$ would change  
from $10^{-10}$ to $10^{-x}$ where $x = 10/ (26 \times 0.15) \approx 2.6$. 
Taking into account that this is a very rough estimate, which may easily
miss one order of magnitude, one cannot exclude
the possibility of observing experimentally and numerically
the unstable flow towards the eventual
critical behavior, even in systems with a moderately small breaking of the
O(5) symmetry at the multicritical point.
\bibitem{footnote-DFP-2d}
The same analysis can be performed in two dimensions: 
the DFP turns out to be stable for $n_1=1$, $n_2=2$ and 
$n_1=2$, $n_2=2$.
\bibitem{footnote1}
The projected SO(5) model  \cite{ZHAHA-99} 
was introduced to overcome some inconsistencies
between the original SO(5) theory and the physics of 
the Mott insulating gap.
\end{references}
\end{document}